\documentclass[aps,prd,twocolumn]{revtex4-2}
\usepackage{hyperref}
\usepackage{amsmath}
\usepackage{amssymb}
\usepackage{graphicx}
\usepackage{xcolor}

\begin{document}

\title{SU(2) gauge theory with fermions on a semi-simple cubic lattice}

\author{Randy Lewis}
\author{Shidsa Pourbakhsh}
\author{Arnab Pradhan}
\author{Lance Siquioco}
\affiliation{Department of Physics and Astronomy, York University, Toronto, Ontario, Canada, M3J 1P3}

\date{June 19, 2026}

\begin{abstract}
A practical Hamiltonian approach to lattice gauge theories would provide access to several
important areas of phenomenology that have been beyond the reach of conventional lattice methods.
Quantum computers seem to be a natural platform for this approach.
With near-term quantum computers in mind,
our work considers a three-dimensional spatial lattice that can host fermions and non-Abelian gauge fields while needing fewer qubits than a simple cubic lattice.
Specifically, the semi-simple cubic (ssc) lattice is obtained by removing half of the gauge links from a standard cubic lattice
in such a way that every vertex becomes trivalent, which streamlines the handling of Gauss's law.
The ssc lattice is topologically equivalent to the triamond lattice but, because the gauge links at each vertex span all three directions, the ssc lattice
can accommodate a local fermion derivative.
The case of staggered fermions with SU(2) gauge fields is presented here.
\end{abstract}

\maketitle

\section{Motivation}

Quantum field theories have wide applicability in several branches of physics, including particle physics where the standard model is a collection of three
intertwined gauge theories.
For decades, lattice methods implemented on classical computers have provided access to strongly coupled gauge theories such as quantum chromodynamics (QCD),
making critical contributions to fundamental progress in this field \cite{Gattringer:2010zz,FlavourLatticeAveragingGroupFLAG:2024oxs}.
Mathematical sign problems have prevented the traditional Monte Carlo approach from addressing some important topics, but
the advent of quantum computers is raising the possibility of overcoming these barriers through the use of Hamiltonian methods
\cite{Bauer:2022hpo,DiMeglio:2023nsa,Halimeh:2025vvp}.

A simple cubic lattice in four-dimensional Euclidean spacetime has been the standard orthonormal framework for most classical lattice computations.
Moving to the Hamiltonian method means the lattice only needs to span the three spatial dimensions (3D), but it also means that Gauss's law must
be actively maintained at each lattice site along with some form of truncation to fit the gauge links onto a finite qubit register
\cite{Byrnes:2005qx,Zohar:2014qma,Zohar:2015hwa,Martinez:2016yna,Raychowdhury:2018tfj,Kaplan:2018vnj,Stryker:2018efp,Raychowdhury:2018osk,Alexandru:2019nsa,Klco:2019evd,Raychowdhury:2019iki,Yang:2020yer,Ji:2020kjk,Davoudi:2020yln,Ciavarella:2021nmj,Atas:2021ext,ARahman:2021ktn,Mazzola:2021hma,VanDamme:2021njp,Ciavarella:2021lel,ARahman:2022tkr,Farrell:2022wyt,Atas:2022dqm,Farrell:2022vyh,Davoudi:2022xmb,Zache:2023cfj,Zache:2023dko,Hayata:2023puo,Muller:2023nnk,Ciavarella:2023mfc,DAndrea:2023qnr,Ebner:2023ixq,Gustafson:2023kvd,Kavaki:2024ijd,Ebner:2024mee,Turro:2024pxu,Calajo:2024qrc,Ciavarella:2024fzw,Carena:2024dzu,Illa:2024kmf,Mathew:2024bed,Gustafson:2024kym,Spagnoli:2024mib,Kadam:2024ifg,Fontana:2024rux,Grabowska:2024emw,Burbano:2024uvn,Lee:2024jnt,Hayata:2024fnh,Gonzalez-Cuadra:2024xul,Ebner:2024qtu,Ciavarella:2024lsp,Halimeh:2024bth,Ballini:2024qmr,Than:2024zaj,Turro:2025sec,Kavaki:2025hcu,Balaji:2025afl,Illa:2025dou,Ciavarella:2025bsg,Jiang:2025ufg,Cobos:2025krn,Joshi:2025pgv,Huie:2025yzn,Ciavarella:2025tdl,Cataldi:2025cyo,Balaji:2025yua,Perez:2025cxl,Yao:2025cxs,Kadam:2025trs,Froland:2025bqf,Chen:2026hnh,Hayata:2026rmv,Modi:2026syn,Rhodes:2026atz,John:2026gut,Webb-Mack:2026bkg,Jha:2026ror,Froland:2026aff}.
For a non-Abelian theory like QCD, the Gauss law conditions require additional care and ultimately additional qubits in the quantum simulation.
For example, an extra set of qubits at each lattice site can keep track of the QCD color sum for the two gauge links arriving from the $\pm x$ directions,
another set of qubits can handle the $\pm y$ sum, and another set the $\pm z$ sum.
Gauss's law insists that these partial sums combine with each other and with the lattice site's fermionic degrees of freedom to produce an overall gauge singlet.

Trivalent vertices are a convenient option for handling Gauss's law
\cite{Raychowdhury:2018tfj,Klco:2019evd,ARahman:2021ktn,Ciavarella:2021lel,ARahman:2022tkr,Hayata:2023puo,Muller:2023nnk,Ebner:2023ixq,Kavaki:2024ijd,Ebner:2024mee,Turro:2024pxu,Kadam:2024ifg,Burbano:2024uvn,Lee:2024jnt,Hayata:2024fnh,Ebner:2024qtu,Turro:2025sec,Kavaki:2025hcu,Illa:2025dou,Cobos:2025krn,Kadam:2025trs,Chen:2026hnh,Hayata:2026rmv,Rhodes:2026atz,John:2026gut}.
In the absence of fermions, a site with only three SU(2) gauge fields has a unique solution for Gauss's law when the gauge field quantum numbers are known.
Fermions can be accommodated with appropriate additional qubits.
The most symmetric 2D lattice having exclusively trivalent vertices is the honeycomb lattice
\cite{Raychowdhury:2018tfj,Hayata:2023puo,Muller:2023nnk,Ebner:2023ixq,Turro:2024pxu,Lee:2024jnt,Ebner:2024qtu,Illa:2025dou,Cobos:2025krn,Chen:2026hnh,Hayata:2026rmv}.
The most symmetric 3D lattice having exclusively trivalent vertices is the triamond lattice \cite{Kavaki:2024ijd,Kavaki:2025hcu},
also called the Laves or $K_4$ lattice \cite{Coxeter,Sunada}.
Other important examples include a single ladder of square plaquettes (which is intermediate between 1D and 2D)
\cite{Klco:2019evd,ARahman:2021ktn,Ciavarella:2021lel,ARahman:2022tkr,Ebner:2024mee,Hayata:2024fnh,Turro:2025sec,John:2026gut} and the 3D hyperhoneycomb lattice \cite{Illa:2025dou}.
In the present work, we add the 3D semi-simple cubic (ssc) lattice \cite{Kuzmin} to this list.

Although the triamond lattice has the largest symmetry group of any 3D trivalent lattice, it presents a complication for fermions.
The three gauge links touching any triamond lattice site always lie in a plane.
This allows for nearest-neighbor fermion derivatives spanning that plane, but any derivative outside the plane would require next-nearest-neighbor terms.
The absence of a derivative beyond the plane prevents construction of a nearest-neighbor 3D Dirac Hamiltonian.
The same situation applies to the hyperhoneycomb lattice \cite{Illa:2025dou}.
In contrast, every site of the ssc lattice has a gauge link in each of the $x,y,z$ directions, thus allowing nearest-neighbor derivatives in all directions.

\begin{figure}
\includegraphics[width=55mm]{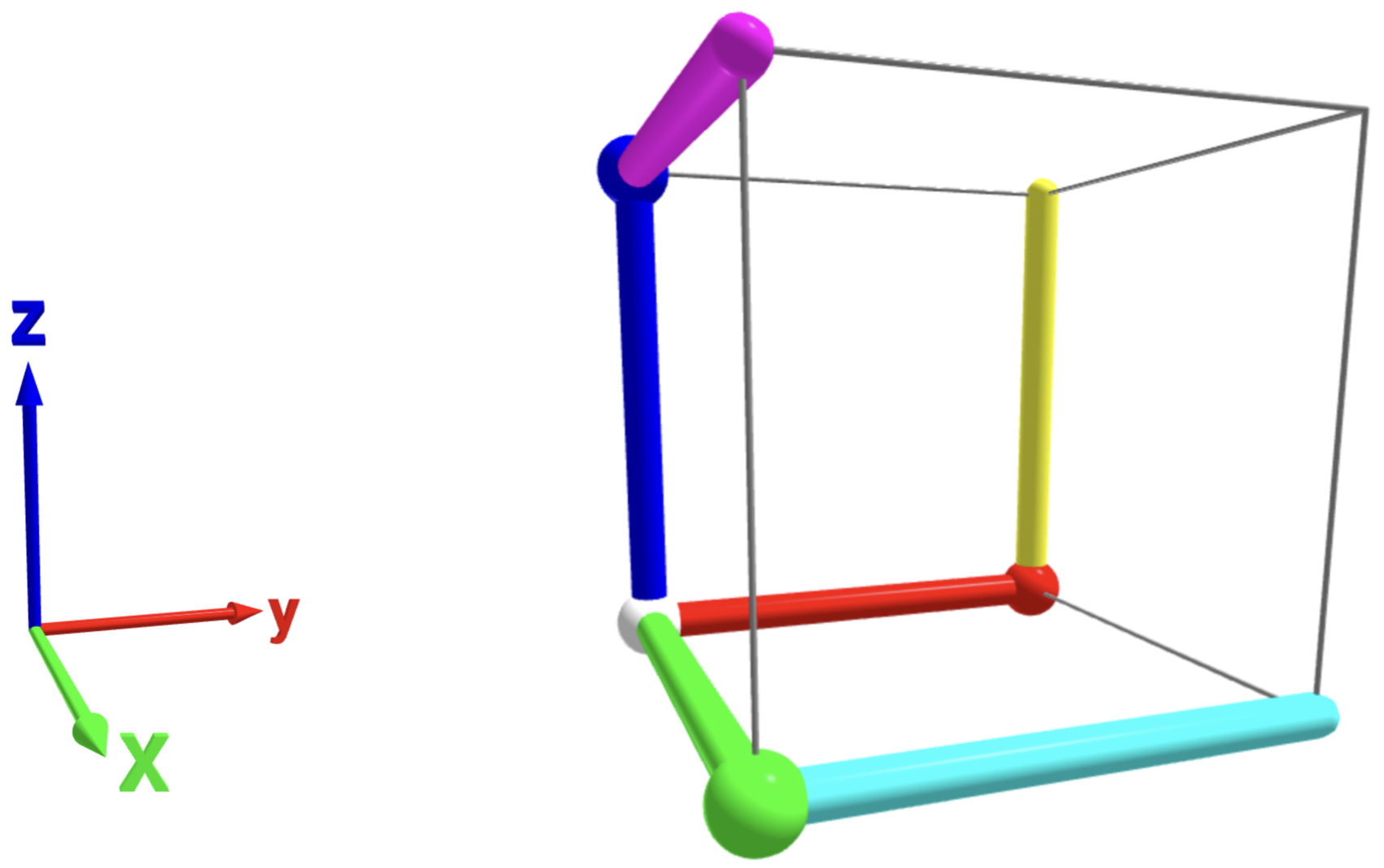}
\caption{The basic building block for the ssc lattice, comprising four lattice sites and six gauge links.
         Colors are not physical but they match those used in discussions of the triamond lattice \cite{Kavaki:2024ijd,Kavaki:2025hcu}.
}\label{fig:ssc1x1}
\end{figure}
\begin{figure}
\includegraphics[width=70mm]{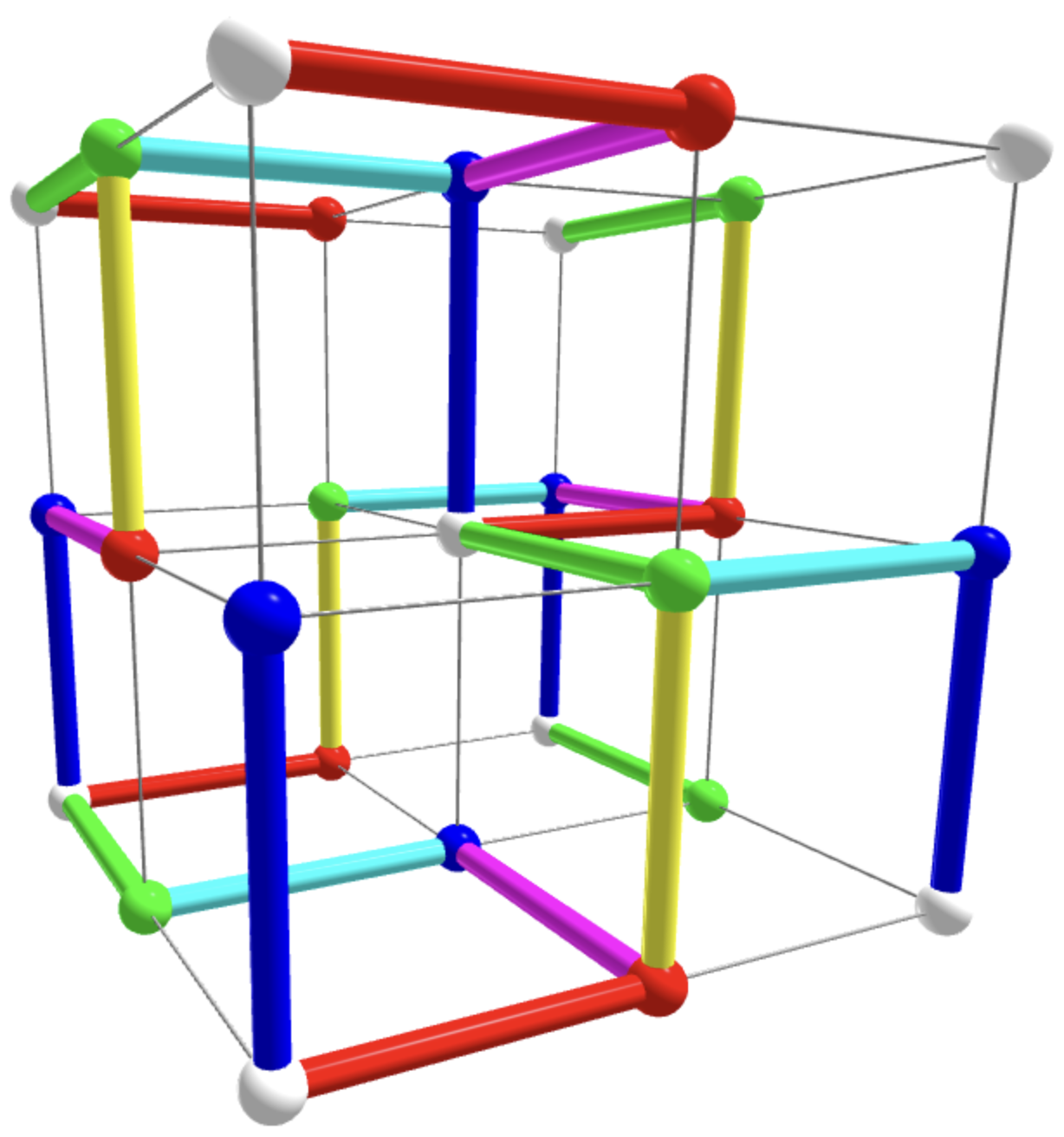}
\caption{A small ssc lattice.
         Colors match those in Fig.~\ref{fig:ssc1x1} and help to show the body-centered cubic pattern.
}\label{fig:ssc2x2}
\end{figure}

The ssc lattice is easily obtained by putting copies of the basic building block from Fig.~\ref{fig:ssc1x1} into a body-centered cubic pattern.
A small ssc lattice is shown in Fig.~\ref{fig:ssc2x2}.
The colors in these diagrams have no physical meaning but are used to make the body-centered cubic pattern easier to see.
(Notice the white sites, for example.)
The colors also allow the reader to make comparisons of the ssc lattice to the triamond lattice because
the same color choices have been used in that literature \cite{Kavaki:2024ijd,Kavaki:2025hcu}.

The ssc lattice keeps the qubit count low by retaining only half of the gauge links from a simple cubic lattice.
The additional economy arising from trivalent vertices means the qubit count is less than half of that for the original cubic lattice.
To incorporate qubit-efficient matter fields, a well-known choice is staggered fermions \cite{Kogut:1974ag,Gattringer:2010zz}.
The staggering procedure retains only select fermion components at each lattice site, in an organized pattern.
The full fermion spinor is still available by combining the degrees of freedom from nearby sites.

Use of an ssc lattice does not specify any particular mapping onto qubits.
With an eye toward keeping the number of qubits small, the present work chooses a scenario where the non-Abelian gauge projections have been summed in advance,
leaving a single quantum number to uniquely define each gauge link during a lattice simulation.

Section~\ref{sec:gauge} of the present work derives the Hamiltonian for SU(2) pure gauge theory on an ssc lattice.
Section~\ref{sec:angmom} discusses the rotational symmetries of an ssc lattice and the connection to angular momentum in a gauge theory.
Section~\ref{sec:theta} shows how the topological theta term, which is very small in QCD but contains a mathematical sign problem for Monte Carlo methods,
can be defined on the ssc lattice.
Section~\ref{sec:fermions} describes our approach to putting staggered fermions onto the ssc lattice.
For the case where SU(2) gauge projections are summed in advance, an explicit mapping onto qubits is explained in Sec.~\ref{sec:mapping}
and applied to the smallest periodic ssc lattice in Sec.~\ref{sec:unitcell}.
Section~\ref{sec:outlook} provides a concluding discussion and outlook.

\section{Pure gauge Hamiltonian}\label{sec:gauge}

Even without any matter fields, SU(2) gauge theory sustains rich nonlinear dynamics among the gauge fields.
The Hamiltonian is conveniently written in terms of chromoelectric and chromomagnetic contributions.
The chromoelectric Hamiltonian is a sum over all gauge links in the lattice \cite{Kogut:1974ag},
\begin{equation}\label{eq:HE}
H_E = \frac{g^2}{2a}\sum_{\rm links}{\rm Tr}\Big(E_x^2(\vec n)+E_y^2(\vec n)+E_z^2(\vec n)\Big) \,,
\end{equation}
where $g$ is the gauge coupling, $a$ is the lattice spacing, $\vec E(\vec n)$ is the chromoelectric field at lattice site $\vec n$,
and the trace is over SU(2) components.
Although $H_E$ has this same form for either a simple cubic lattice or an ssc lattice, the sum in Eq.~(\ref{eq:HE})
has only half the number of terms in the ssc case
because the ssc lattice has half the links of a simple cubic lattice.

\begin{figure}
\includegraphics[width=70mm]{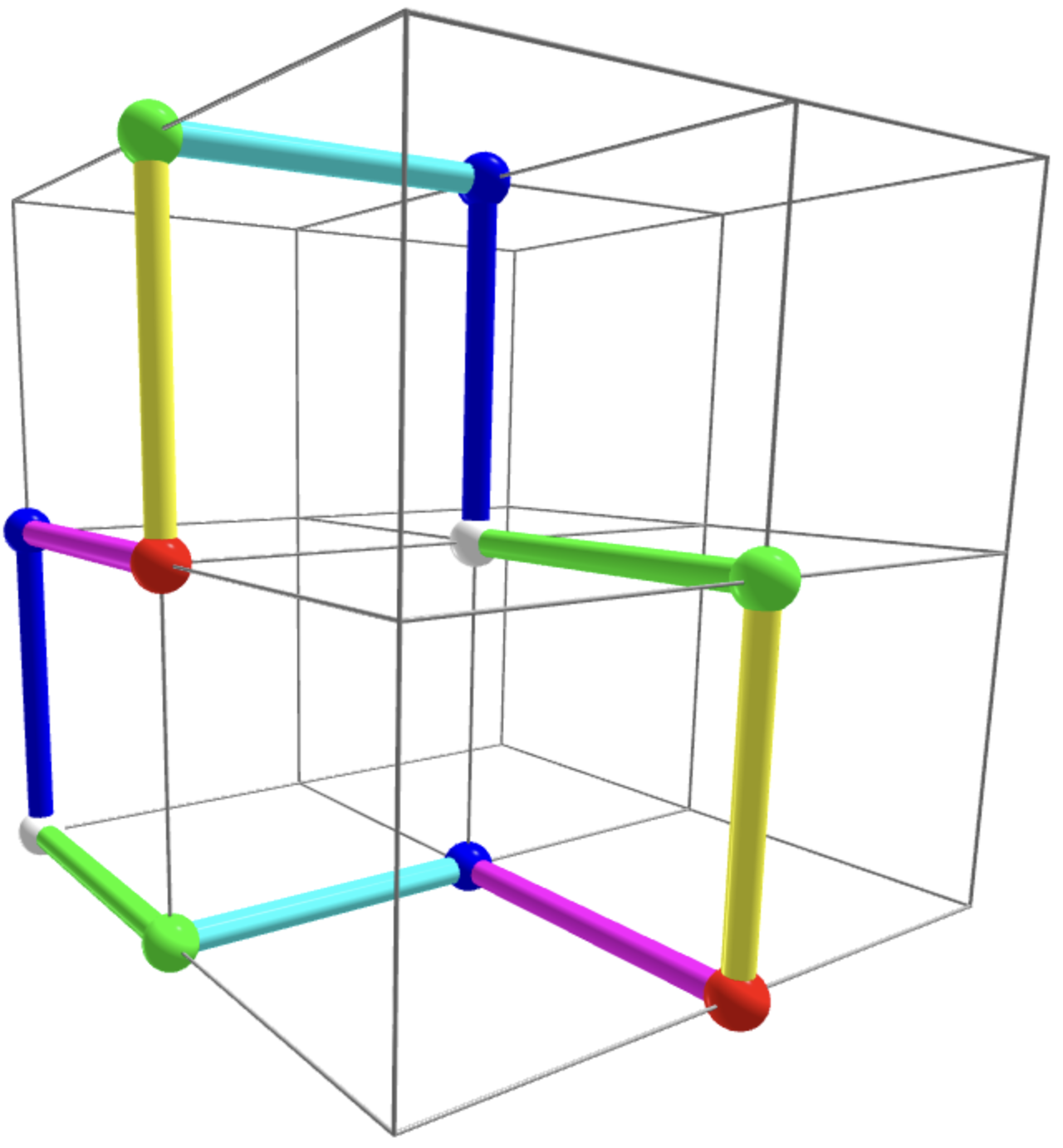}
\caption{One plaquette from the ssc lattice.
         All other plaquettes are obtained from this one via spatial rotations and translations.
}\label{fig:plaquette}
\end{figure}

The chromomagnetic Hamiltonian is written as a sum over elementary plaquettes \cite{Kogut:1974ag}.
These would be four-sided squares in a simple cubic lattice.
Because half of those links are absent in the ssc lattice, we should expect the plaquettes to be larger.
Indeed, an elementary plaquette is a ten-sided path as shown in Fig.~\ref{fig:plaquette}.
All plaquettes have this exact shape and are found with six different orientations throughout the lattice.
In terms of our color scheme, notice that the plaquette in Fig.~\ref{fig:plaquette} uses each color twice except red, which does not appear at all.
Every nonred plaquette has the same orientation as Fig.~\ref{fig:plaquette} on the lattice.
The other orientations are nongreen, nonblue, noncyan, nonmagenta and nonyellow.
Topologically, the six plaquette types are identical to those of the triamond lattice \cite{Kavaki:2024ijd,Kavaki:2025hcu}.

To derive the chromomagnetic Hamiltonian $H_B$, we can begin with the nonred plaquette of Fig.~\ref{fig:plaquette},
\begin{equation}
P_{\bar r}(\vec n) = {\rm Tr}\left(U_y(\vec n)V_z(\vec n+\hat y)V_y^\dagger(\vec n+\hat z)V_z^\dagger(\vec n)\right) \,,
\end{equation}
where $U_y(\vec n)$ is the cyan gauge link at the bottom of the figure and the other gauge links are written as three-sided staples,
\begin{eqnarray}
V_z(\vec n+\hat y) &=& U_x(\vec n+\hat y)U_z(\vec n+\hat x+\hat y)U_x^\dagger(\vec n+\hat y+\hat z) \,,~~~\\
V_y^\dagger(\vec n+\hat z) &=& U_z(\vec n+\hat y+\hat z)U_y^\dagger(\vec n+2\hat z)U_z^\dagger(\vec n+\hat z) \,, \\
V_z^\dagger(\vec n) &=& U_x^\dagger(\vec n-\hat x+\hat z)U_z^\dagger(\vec n-\hat x)U_x(\vec n-\hat x) \,.
\end{eqnarray}
A link variable depends on the underlying gauge field $\vec A(\vec n)$ as
\begin{equation}
U_j(\vec n+\hat k) = e^{iagA_j(\vec n+a\hat k)} \,,
\end{equation}
which allows an expansion in powers of $a$.
In this way, the nonred plaquette is found to be
\begin{equation}
P_{\bar r}(\vec n) = {\rm Tr}(I) - 2g^2a^4{\rm Tr}\left(\left(F_{xz}(\vec n)+F_{yz}(\vec n)\right)^2\right) + O(a^6),
\end{equation}
where ${\rm Tr}(I)=2$ for SU(2) and the field strength tensor is
\begin{equation}
F_{jk}(\vec n) = \partial_jA_k(\vec n)-\partial_kA_j(\vec n) + ig\left[A_j(\vec n),A_k(\vec n)\right] \,.
\end{equation}
The noncyan plaquette differs from the nonred case only in the sign of the cross term,
\begin{equation}
P_{\bar c}(\vec n) = {\rm Tr}(I) - 2g^2a^4{\rm Tr}\left(\left(F_{xz}(\vec n)-F_{yz}(\vec n)\right)^2\right) + O(a^6).
\end{equation}
The nongreen and nonmagenta plaquettes also form a pair,
\begin{equation}
P_{\bar g,\bar m}(\vec n) = {\rm Tr}(I) - 2g^2a^4{\rm Tr}\left(\left(F_{zy}(\vec n)\pm F_{xy}(\vec n)\right)^2\right) + O(a^6),
\end{equation}
as do the nonblue and nonyellow plaquettes,
\begin{equation}
P_{\bar b,\bar y}(\vec n) = {\rm Tr}(I) - 2g^2a^4{\rm Tr}\left(\left(F_{yx}(\vec n)\pm F_{zx}(\vec n)\right)^2\right) + O(a^6).
\end{equation}

Every simple cubic lattice with periodic boundary conditions has three plaquettes per lattice site, but the ssc lattice has half that number.
Therefore the chromomagnetic Hamiltonian that is already known from the continuum theory,
\begin{equation}
H_B = a^3{\rm Tr}\left(F_{xy}^2(\vec n)+F_{yz}^2(\vec n)+F_{zx}^2(\vec n)\right) \,,
\end{equation}
is obtained from the following sum over ssc plaquettes,
\begin{equation}\label{eq:HB}
H_B = \frac{1}{2ag^2}\sum_{\rm plaqs}\left({\rm Tr}(I)-P_k(\vec n)\right) \,.
\end{equation}
To be explicit, the sum runs over all plaquettes on the ssc lattice, including all locations $\vec n$ and all orientations
$k\in\{\bar r,\bar g,\bar b,\bar c,\bar m,\bar y\}$, keeping in mind that
the total number of plaquettes on a periodic ssc lattice having $N$ sites is $3N/2$, not $6N$.

\section{Rotational symmetries and angular momentum}\label{sec:angmom}

The simple cubic lattice has been famously successful for QCD phenomenology, including quantized angular momentum,
despite having only a discrete subgroup of continuous rotational symmetry.
Here we show that the ssc lattice likewise retains a subgroup that is sufficient for the phenomenology of angular momentum.

To begin, imagine rotating the ssc lattice shown in Fig.~\ref{fig:ssc2x2} around either the $x$, $y$ or $z$ axis by angle $\pi$.
The lattice is not invariant.
However, the lattice is invariant under a $\pi$ rotation around $z$ followed by a translation of the lattice by distance $a$
in the $y$ direction.
(Some colors get interchanged, but remember that colors are unphysical.)
The lattice is similarly invariant under $y$ rotation accompanied by $x$ translation,
and also under $x$ rotation accompanied by $z$ translation.
These are called screw symmetries \cite{Kuzmin}.
All three screw symmetries are present at each lattice site.

Besides the 2-fold symmetry axes just discussed, the ssc lattice also has a 3-fold rotation axis at every lattice site.
For example, consider the white site at the center of Fig.~\ref{fig:ssc2x2}.
A 3-fold rotation about the appropriate body-diagonal axis will move the attached red link to the green link's location, green to blue, and blue to red.
The entire lattice is invariant under this $2\pi/3$ rotation.
For every site on the lattice, there is a rotation axis parallel to one of the four body diagonals.

To summarize, the rotation group for the ssc lattice is $I2_13$ in Hermann–Mauguin notation \cite{Kuzmin},
where the $I$ means body-centered cubic,
the integers represent the 2-fold and 3-fold symmetries, and the subscript indicates screw symmetry.
(As an aside, we mention that $I2_13$ is a subgroup of the triamond lattice symmetry group $I4_132$ \cite{Kuzmin},
though we will not need this triamond connection in the present work.)
The screw symmetries of the ssc lattice provide access to eigenstates of a combined linear and angular momentum.
Here we choose to set the linear momentum to zero so we can focus on angular momentum.

To create a state with zero linear momentum,
consider having a single nonred plaquette activated while all other gauge links are set to zero.
The democratic superposition of all possible locations for this nonred plaquette is a zero-momentum state,
$\sum_{\vec n} P_{\bar r}(\vec n)$.
Five other zero-momentum states are obtained from the other orientations (nongreen, nonblue, etc.).
Summing all six of the zero-momentum states results in a state that is invariant under the entire $I2_13$ symmetry,
\begin{equation}\label{eq:scalar}
\sum_{\vec n}\Big( P_{\bar r}(\vec n) + P_{\bar g}(\vec n) + P_{\bar b}(\vec n) + P_{\bar c}(\vec n) + P_{\bar m}(\vec n) + P_{\bar y}(\vec n) \Big) \,,
\end{equation}
which corresponds to zero angular momentum.
A spin-zero glueball in the SU(2) theory will couple to this scalar state.

\begin{figure}
\includegraphics[width=80mm]{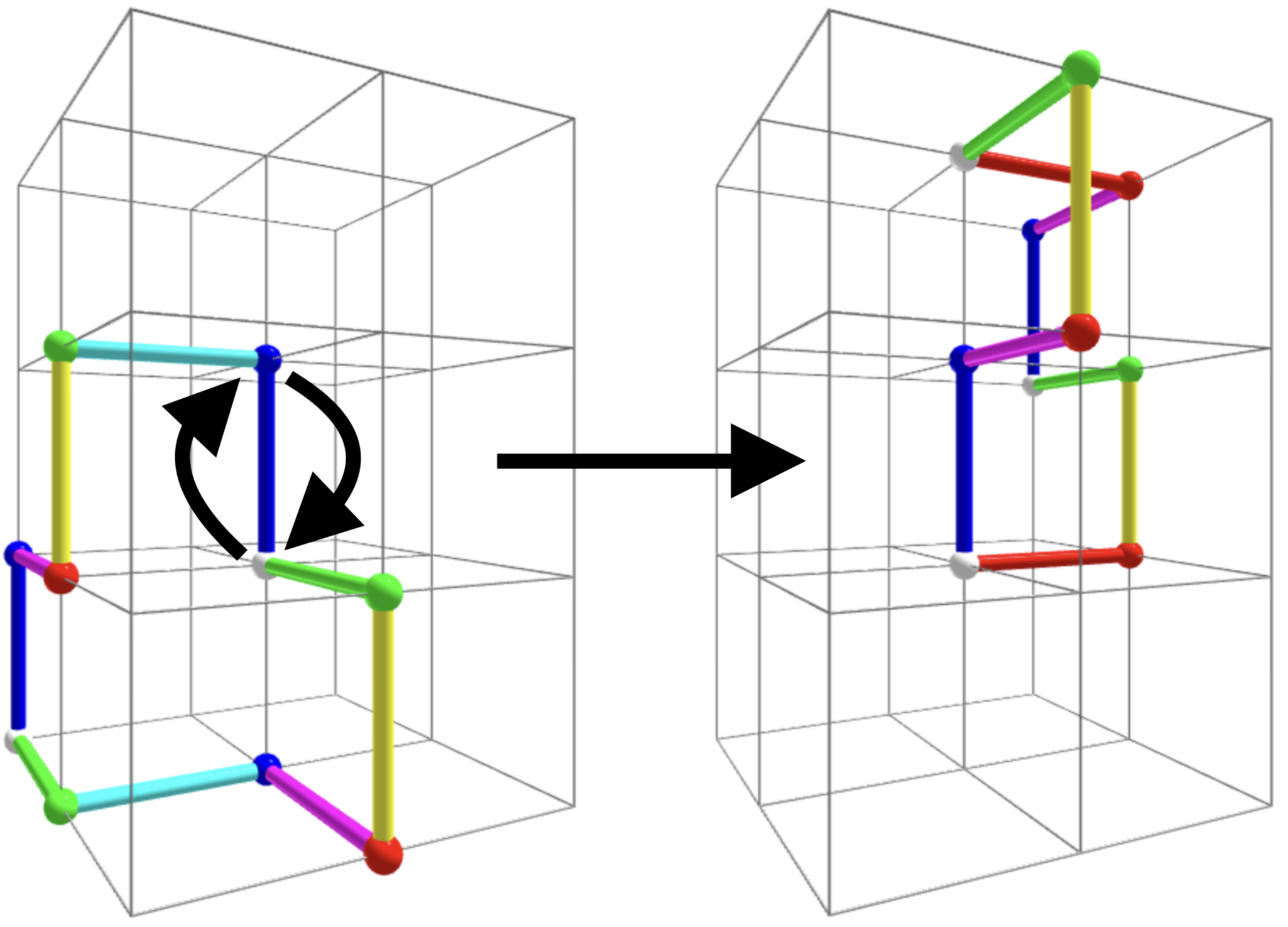}
\caption{Application of screw symmetry to a single plaquette.
         Rotation around the $x$ axis (at the white site) followed by translation in the $z$ direction converts
         a nonred plaquette into a noncyan plaquette.
         Here, that two-step process is shown in one step by rotating around the midpoint of a blue link.
}\label{fig:rotplaq}
\end{figure}

A different linear combination of plaquettes should produce a spin-one state.
To find it, consider various rotations of the nonred plaquette in Fig.~\ref{fig:plaquette}.
Rotation by angle $\pi$ around a vertical axis leaves it invariant (up to translations),
but a $\pi$ rotation around either the $x$ or $y$ axis turns the plaquette upside down,
meaning it becomes a noncyan plaquette as shown in Fig.~\ref{fig:rotplaq}.
This indicates that the difference of nonred and noncyan,
\begin{equation}\label{eq:vector1}
\sum_{\vec n}\left( P_{\bar r}(\vec n) - P_{\bar c}(\vec n) \right) \,,
\end{equation}
has the rotation properties of a vector aligned with the $z$ axis.
Similarly nongreen minus nonmagenta,
\begin{equation}
\sum_{\vec n}\left( P_{\bar g}(\vec n) - P_{\bar m}(\vec n) \right) \,,
\end{equation}
is a vector aligned with the $y$ axis while nonblue minus nonyellow,
\begin{equation}\label{eq:vector3}
\sum_{\vec n}\left( P_{\bar b}(\vec n) - P_{\bar y}(\vec n) \right) \,,
\end{equation}
is a vector aligned with the $x$ axis.
A spin-one glueball in the SU(2) theory will couple to these vector states.

Higher spins can be handled as well, and for that we find it useful to invoke the group theory of conjugacy classes and irreducible representations applied to the tetrahedral group.
Appendix~\ref{app:tetrahedral} provides a brief introduction for readers who might be interested.

\section{The topological theta term}\label{sec:theta}

For most calculations, $H_E$ and $H_B$ from Sec.~\ref{sec:gauge} are the only pure gauge terms required.
However, there is an additional term called the theta term \cite{Bonanno:2026zzf}, which arises from the following spacetime Lagrangian density,
\begin{equation}
{\cal L}(\vec n) = \frac{g^2\theta}{32\pi^2}\epsilon^{\mu\nu\lambda\rho}{\rm Tr}\left(F_{\mu\nu}(\vec n)F_{\lambda\rho}(\vec n)\right) \,.
\end{equation}
This term is neglected for most QCD calculations because experiments require $\theta<10^{-10}$ \cite{Abel:2020pzs},
though the reason for such a small numerical value is an ongoing discussion.
Larger values of $\theta$ can be of interest in non-QCD theories, including SU(2).

Lattice Monte Carlo calculations encounter a sign problem when the theta term is included, but the sign problem is avoided
in the Hamiltonian framework as discussed for a simple cubic lattice in Ref.~\cite{Kan:2021nyu}.
Here we derive the expression for the theta term on an ssc lattice.

To begin, recall that $F_{\mu\nu}$ is the chromoelectric field when either $\mu$ or $\nu$ is the time direction.
Therefore we can write
\begin{equation}
{\cal L}(\vec n) = \frac{g^2\theta}{8\pi^2a^2}\epsilon^{ijk}{\rm Tr}\left(E_i(\vec n)F_{jk}(\vec n)\right) \,.
\end{equation}
On a simple cubic lattice, the untraced product of gauge links around a plaquette is
\begin{eqnarray}
U_{jk}(\vec n) &=& U_j(\vec n)U_k(\vec n+\hat j)U_j^\dagger(\vec n+\hat k)U_k^\dagger(\vec n) \\
               &=& I + ia^2gF_{jk}(\vec n) + O(a^3) \,.
\end{eqnarray}
Therefore the field strength tensor is
\begin{equation}
F_{jk}(\vec n) = -\frac{i}{2a^2g}\left(U_{jk}(\vec n)-U_{jk}^\dagger(\vec n)\right) \,,
\end{equation}
which leads to the Hamiltonian
\begin{equation}
H_\text{cubic} = -\frac{ig^2\theta}{16\pi^2a}\epsilon^{ijk}\sum_{\vec n}{\rm Tr}\left(E_i(\vec n)\left(U_{jk}(\vec n)-U_{jk}^\dagger(\vec n)\right)\right)
\end{equation}
as obtained in Ref.~\cite{Kan:2021nyu}.

On an ssc lattice, the untraced plaquettes are labeled by their orientations (nonred, noncyan, etc).
The expressions are
\begin{eqnarray}
U_{\bar r}(\vec n) &=& I + 2ia^2g\left(F_{xz}(\vec n) + F_{yz}(\vec n)\right) + O(a^3) \,, \\
U_{\bar c}(\vec n) &=& I + 2ia^2g\left(F_{xz}(\vec n) - F_{yz}(\vec n)\right) + O(a^3) \,, \\
U_{\bar g}(\vec n) &=& I + 2ia^2g\left(F_{zy}(\vec n) + F_{xy}(\vec n)\right) + O(a^3) \,, \\
U_{\bar m}(\vec n) &=& I + 2ia^2g\left(F_{zy}(\vec n) - F_{xy}(\vec n)\right) + O(a^3) \,, \\
U_{\bar b}(\vec n) &=& I + 2ia^2g\left(F_{yx}(\vec n) + F_{zx}(\vec n)\right) + O(a^3) \,, \\
U_{\bar y}(\vec n) &=& I + 2ia^2g\left(F_{yx}(\vec n) - F_{zx}(\vec n)\right) + O(a^3) \,.~~~~~
\end{eqnarray}
Simple differences provide the field strength components needed for the theta term,
\begin{eqnarray}
F_{yz}(\vec n) &=& -\frac{i}{4a^2g}\left(U_{\bar r}(\vec n)-U_{\bar c}(\vec n)\right) \,, \\
F_{xy}(\vec n) &=& -\frac{i}{4a^2g}\left(U_{\bar g}(\vec n)-U_{\bar m}(\vec n)\right) \,, \\
F_{zx}(\vec n) &=& -\frac{i}{4a^2g}\left(U_{\bar b}(\vec n)-U_{\bar y}(\vec n)\right) \,,
\end{eqnarray}
giving the Hamiltonian
\begin{eqnarray}
H_\theta &=& -\frac{ig^2\theta}{16\pi^2a}\sum_{\vec n}{\rm Tr}\Big(E_x(\vec n)\left(U_{\bar r}(\vec n)-U_{\bar c}(\vec n)\right) \nonumber \\
&& + E_y(\vec n)\left(U_{\bar b}(\vec n)-U_{\bar y}(\vec n)\right) \nonumber \\
&& + E_z(\vec n)\left(U_{\bar g}(\vec n)-U_{\bar m}(\vec n)\right)\Big) \,. \label{eq:Htheta}
\end{eqnarray}
The sum $H_E+H_B+H_\theta$ is the complete Hamiltonian for a pure gauge theory.

\section{Staggered fermion Hamiltonian}\label{sec:fermions}

Naively discretizing Dirac fermions onto a lattice results in a well-known doubling problem.
Staggering is one of the ways to resolve it \cite{Gattringer:2010zz,Kogut:1974ag}.
(For ease of notation, we are choosing $a=1$ from now on.)
To begin, recall the naive Dirac Hamiltonian on a lattice,
\begin{equation}\label{eq:naiveH}
H = \sum_{\vec n}\sum_{\vec n^\prime}\bar\Psi(\vec n)\,\left(i\gamma_j\Delta_j(\vec n,\vec n^\prime)+m\delta_{\vec n,\vec n^\prime}\right)\,\Psi(\vec n^\prime) \,,
\end{equation}
where $\vec n$ and $\vec n^\prime$ are two lattice sites, $j$ is summed over spatial directions $x,y,z$, and the covariant derivative is
\begin{equation}\label{eq:Delta}
\Delta_j(\vec n,\vec n^\prime) = \frac{1}{2}\left(\delta_{\vec n+\hat j,\vec n^\prime}W_j(\vec n)
                                 -\delta_{\vec n-\hat j,\vec n^\prime}W_j^\dagger(\vec n-\hat j)\right) \,.
\end{equation}
For a simple cubic lattice, the minimal choice uses single gauge fields: $W_j(\vec n)=U_j(\vec n)$ and $W_j^\dagger(\vec n-\hat j)=U_j^\dagger(\vec n-\hat j)$
but longer gauge paths represented by products of $U$ factors can be used instead, provided the endpoints do not change.
Monte Carlo lattice calculations often use a superposition of several paths to enhance overlap with the phenomenology of interest.

\begin{figure}
\includegraphics[width=35mm]{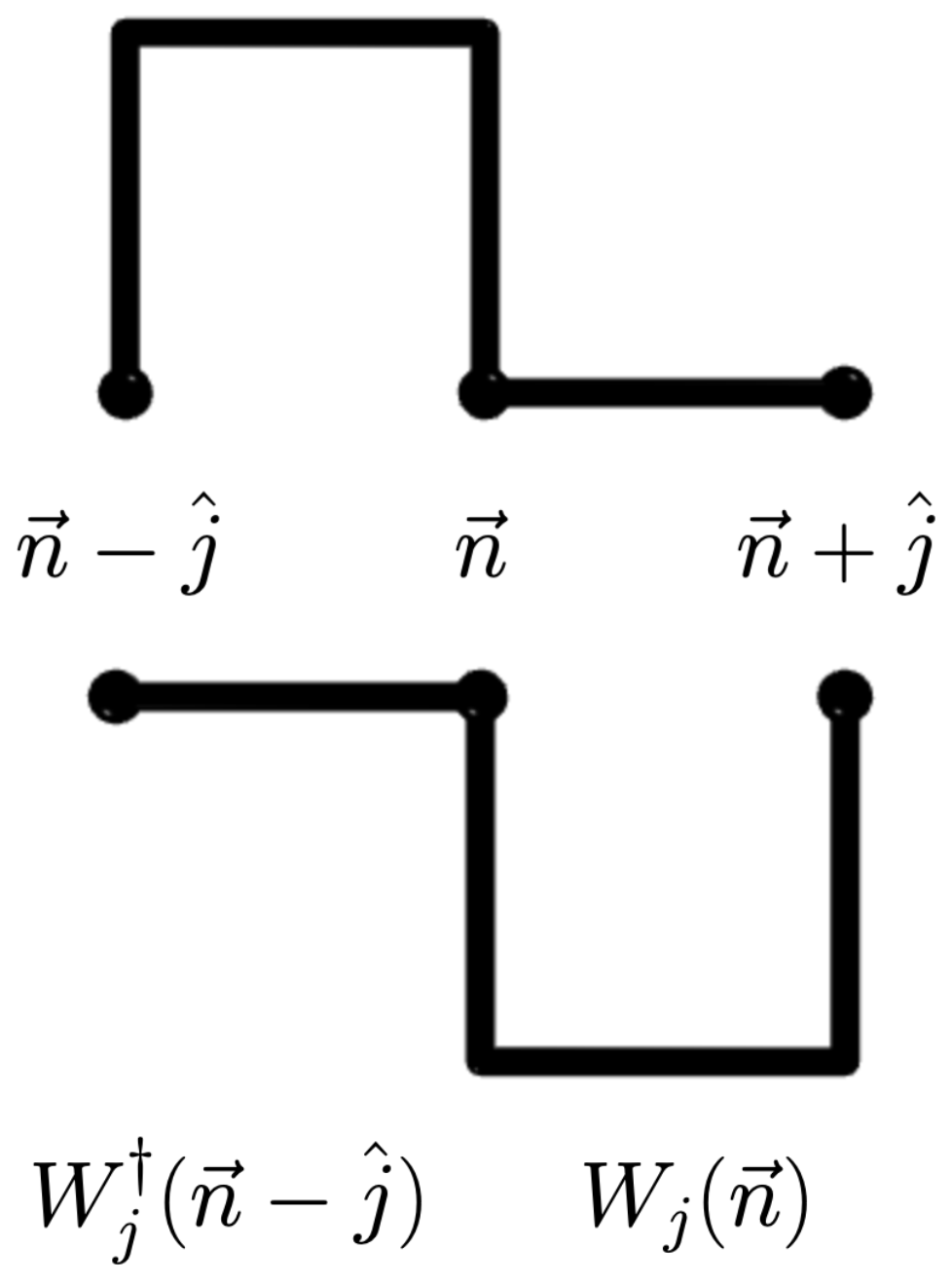}
\caption{The two options for $W_j(\vec n)$ and $W_j^\dagger(\vec n-\hat j)$ of Eq.~(\ref{eq:Delta}).
         Some ssc lattice sites have a forward single link and a backward three-link staple.
         The other sites have a forward three-link staple and a backward single link.
}\label{fig:W}
\end{figure}

For an ssc lattice, 
the minimal choice is for either $W_j(\vec n)$ or $W_j^\dagger(\vec n-\hat j)$ to be a single gauge link but, since the ssc lattice has no single gauge link
in the opposite direction, the other $W$ factor is a three-link staple.
These two situations, which account for every site on an ssc lattice, are displayed in Fig.~\ref{fig:W}.

The staggered ssc Hamiltonian will be obtained with minor adjustments to the method provided in Ref.~\cite{Catterall:2025vrx}.
Step 1 in the staggering procedure is to apply the following transformation to the fermion field of Eq.~(\ref{eq:naiveH}),
\begin{equation}\label{eq:alpha}
\Psi(\vec n) = (i\gamma^0\gamma^1)^{n_x}(i\gamma^0\gamma^2)^{n_y}(i\gamma^0\gamma^3)^{n_z}\chi(\vec n) \,.
\end{equation}
Notice that the transformation has different powers on the Dirac gamma matrices depending on whether the lattice site is even or odd in each direction,
since a site location is $\vec n=(n_x,n_y,n_z)$.
Step 2 in the staggering procedure is to define
\begin{equation}
\chi_\pm = \frac{1}{2}(1\pm\gamma^0)\chi \,.
\end{equation}
This results in a Hamiltonian where $\chi_+$ and $\chi_-$ are decoupled from each other.
Step 3 is to retain only $\chi_+$ (and we will call it simply $\chi$ from now on).
The final fermionic Hamiltonian has mass and kinetic terms as follows,
\begin{eqnarray}
H &=& H_m + H_k \,, \\
H_m &=& \sum_{\vec x}(-1)^{n_x+n_y+n_z}m\chi^\dagger(\vec n)\chi(\vec n) \,, \label{eq:Hm} \\
H_k &=& \sum_{\vec n,\vec n^\prime}\chi^\dagger(\vec n)\Big(\Sigma_x(\vec n,\vec n^\prime)+(-1)^{n_x}\Sigma_y(\vec n,\vec n^\prime) \nonumber \\
    & & +(-1)^{n_x+n_y}\Sigma_z(\vec n,\vec n^\prime)\Big)\chi(\vec n^\prime) + h.c. \,, \label{eq:Hk}
\end{eqnarray}
with
\begin{equation}
\Sigma_j(\vec n,\vec n^\prime) = \frac{1}{2}\left(\delta_{\vec n+\hat j,\vec n^\prime}W_j(\vec n)+\delta_{\vec n-\hat j,\vec n^\prime}W_j^\dagger(\vec n-\hat j)\right) \,.
\end{equation}
Notice that $\Sigma_j(\vec n,\vec n^\prime)$ is a sum of terms whereas $\Delta_j(\vec n,\vec n^\prime)$ was a difference of terms.

An important result of staggering is evident in this fermionic Hamiltonian.
The transformation has removed all explicit Dirac gamma matrices, so the four components in $\chi$ do not mix with each other.
This means we can define a reduced staggered theory by treating $\chi$ as a simple scalar with one component instead of four.
The doubler problem of naive fermions is now absent, and the complete fermion spinor can still be constructed by combining the fermion components from
neighboring lattice sites \cite{Catterall:2025vrx}.

In the low-momentum limit, the staggered fields residing at even sites (meaning $n_x+n_y+n_z$ is even)
correspond to fermion degrees of freedom, while the odd sites correspond to antifermions.
See Appendix~\ref{app:fermionsites} for details.
Because of this, the following sections will refer to even and odd sites as quark and antiquark sites respectively.

Since $H_k$ is linear in both $\chi^\dagger(\vec n)$ and $\chi(\vec n\pm\hat j)$, the fundamental role of $H_k$ is to create or
annihilate a quark-antiquark pair at neighboring lattice sites.
Movement of a fermion across the lattice arises from repeated application of $H_k$.
For example, suppose a quark is present at site $\vec n$ while sites $\vec n+\hat x$ and $\vec n+2\hat x$ are both vacant.
A first application of $H_k$ can create an antiquark at $\vec n+\hat x$ and a quark at $\vec n+2\hat x$, and then a second application of $H_k$
can annihilate the quark at $\vec n$ and the antiquark at $\vec n+\hat x$.
The net result is movement of the original quark from $\vec n$ to $\vec n+2\hat x$, automatically
including the appropriate changes to gauge fields at every step so that Gauss's law is maintained.

\section{An efficient qubit mapping}\label{sec:mapping}

The complete Hamiltonian for SU(2) gauge theory is
\begin{equation}
H_{\rm total} = H_E + H_B + H_\theta + H_m + H_k
\end{equation}
with the individual terms given in Eqs.~(\ref{eq:HE}), (\ref{eq:HB}), (\ref{eq:Htheta}), (\ref{eq:Hm}) and (\ref{eq:Hk}).
Mapping this Hamiltonian onto a qubit register can be done in several ways, each with its own set of advantages.
The example provided in this section focuses on gauge-singlet states, which reduces the number of qubits by eliminating most of the unphysical
non-singlet states that would otherwise be present.

In general, basis states for a gauge link in the SU(2) theory can be expressed as $\left|j,m_L,m_R\right>$ where $j\in\{0,\frac{1}{2},1,\frac{3}{2},\ldots\}$
is the SU(2) quantum number having projection $m_L$ (along a chosen SU(2) direction) at one end and $m_R$ at the other end \cite{Byrnes:2005qx}.
The values of $m_L$ and $m_R$ will combine with adjoining fermions and/or gauge links to ensure that Gauss's law is preserved at every lattice site.
To build a mapping containing only Gauss-preserving states at each site, we can work directly with quantities that have already been summed over $m_L$ and $m_R$,
leaving $j$ as the only parameter for a gauge link.
For any chosen $j_{\rm max}$, the range $0\leq j\leq j_{\rm max}$ can be handled by a collection of qubits or a single qudit.

At any site of an ssc lattice, there are three gauge links arriving, one along the $x$ axis, one along $y$, and one along $z$ with quantum numbers
$j_x$, $j_y$ and $j_z$ respectively.
There might also be one or two staggered fermions at the site but more than two would violate Fermi-Dirac statistics.
This allows four basis states at the site:
option (a) has no fermions present and the three gauge links uniquely obey Gauss’s law,
option (b) has two fermions present as an SU(2) singlet and the three gauge links uniquely obey Gauss’s law,
option (c) has one fermion present and it combines with $j_z$ to give $j_z+\frac{1}{2}$,
and option (d) has one fermion present and it combines with $j_z$ to give $j_z-\frac{1}{2}$.

The $j$ values from the three gauge fields are already enough to indicate whether a given site is within \{(a),(b)\} or within \{(c),(d)\}.
This means just one additional qubit is sufficient to specify any of the four cases uniquely.

For a minimal explicit example, consider truncating all gauge links at $j_{\rm max}=\frac{1}{2}$ and neglecting the
topological theta term.
Our usage of the vertex qubit is detailed in Table~\ref{tab:vertex}.
In what follows, our Pauli basis obeys $Z\left|0\right>=\left|0\right>$ and $Z\left|1\right>=-\left|1\right>$.

\begin{table}
\caption{Roles of the vertex qubit for the case of $j_{\rm max}=\frac{1}{2}$ truncation on each gauge link.
         The total quantum number for the $x$-axis and $y$-axis gauge links is $j_{xy}$.
         The total quantum number for the $z$-axis gauge link and the fermion is $j_{zv}$.
        }\label{tab:vertex}
\begin{ruledtabular}
\begin{tabular}{cccc}
type of & number of & \multicolumn{2}{c}{basis states of vertex qubit} \\
site    & $j=\frac{1}{2}$ links & $\left|0\right>$ & $\left|1\right>$ \\
\hline
quark & 0 & no quarks & two quarks \\
      & 1 & one quark & unphysical \\
      & 2 & no quarks & two quarks \\
      & 3 & $j_{xy}=j_{zv}=0$ & $j_{xy}=j_{zv}=1$ \\
\hline
antiquark & 0 & two antiquarks & no antiquarks \\
          & 1 & one antiquark  & unphysical \\
          & 2 & two antiquarks & no antiquarks \\
          & 3 & $j_{xy}=j_{zv}=0$ & $j_{xy}=j_{zv}=1$
\end{tabular}
\end{ruledtabular}
\end{table}

The chromoelectric term in the Hamiltonian is a sum that contributes $\frac{1}{2}g^2j(j+1)$ for each gauge link on the lattice,
\begin{equation}
H_E = \frac{3g^2}{8}\sum_{n={\rm link}}\left(\frac{1-Z_n}{2}\right) \,.
\end{equation}

The chromomagnetic term in the Hamiltonian flips the $j$ value of the ten links in a plaquette and multiplies by a factor of $-\frac{1}{2}g^{-2}(i\sqrt{2})^{-c}$.
The integer $c$ is the number of sites in the 10-sided plaquette where the pair of plaquette links have unequal $j$ values.
Therefore $0\leq c\leq10$ and the Hamiltonian is
\begin{equation}
H_B = -\frac{1}{2g^2}\sum_{\rm plaqs}\prod_{n=1}^{10}X_n\left(1-\frac{(\sqrt{2}+i)(1-Z_nZ_{n-1})}{2\sqrt{2}}\right) \,.
      \label{eq:HBqubit}
\end{equation}
Note that $n=0$ and $n=10$ point to the same gauge link.
Note also that $c$ is always an even integer so the multiplicative factor (already built into Eq.~(\ref{eq:HBqubit})) is always real.

The mass term in the Hamiltonian adds a mass $m$ for each quark present at a quark site,
and it subtracts a mass $m$ for each antiquark that is absent from an antiquark site,
\begin{equation}
H_m = \sum_{\rm sites}\frac{m}{2}(-1)^p(2-Z_v-Z_xZ_yZ_zZ_v) \,,
\end{equation}
where $Z_v$ acts on the vertex qubit and the other $Z$ operators act on their respective gauge links at this lattice site.
The exponent $p$ is 0 for a quark site and 1 for an antiquark site.

Each kinetic term in the Hamiltonian creates or annihilates a quark-antiquark pair at neighboring lattice sites,
and it flips the $j$ value of the link (or path of three links) connecting those sites.
The kinetic terms in the $z$ direction are
\begin{eqnarray}
\left(H_k\right)_z &=& \sum_{z~{\rm links}}(-1)^{n_x+n_y}D^\dagger_{\vec n}X_zD_{\vec n+\hat z} \,, \label{eq:Hkz} \\
D &=& \frac{1}{4}(1+Z_v)(1+X_v-(1-X_v)Z_xZ_yZ_z) \,, ~~~~~
\end{eqnarray}
where $(-1)^{n_x+n_y}$ is the staggering phase, and $X_z$ acts either on a single gauge link or a path of three gauge links as appropriate.
The kinetic terms in the $x$ and $y$ directions are very similar to the $z$ direction, but
they need to be re-expressed in the $z$ basis which makes their final forms less compact.
Details are provided in Appendix~\ref{app:morekinetic}.
One of the key advantages of using an ssc lattice is that trivalent vertices have reduced this type of complication to a minimum.

\section{The periodic unit cell}\label{sec:unitcell}

\begin{figure}
\includegraphics[width=70mm]{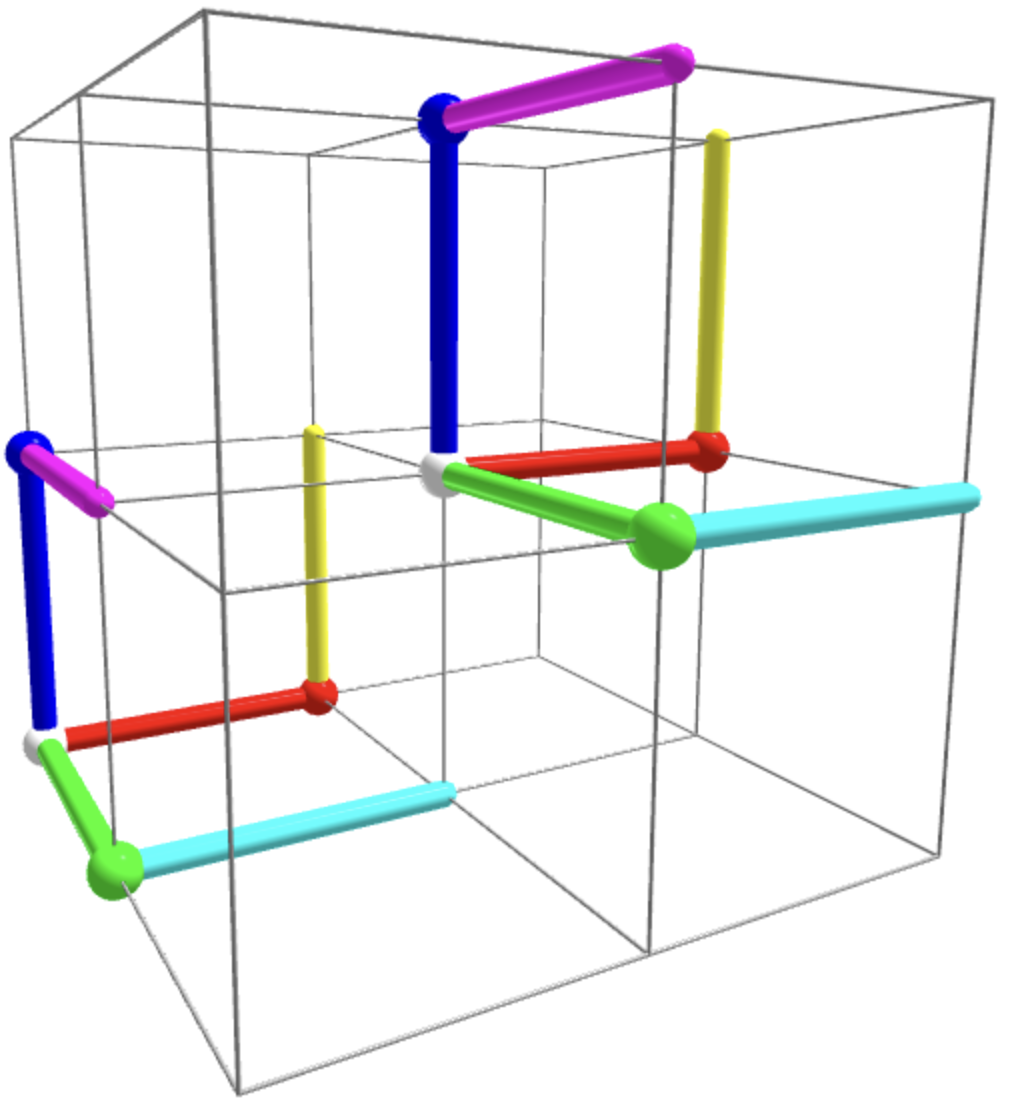}
\caption{A unit cell of the ssc lattice.
         Stacking unit cells together creates a larger ssc lattice.
}\label{fig:unitcell}
\end{figure}

An arbitrarily large ssc lattice can be created by filling space with side-by-side 3D unit cells.
The unit cell is not Fig.~\ref{fig:ssc1x1}, which is the basic building block of volume $a^3$.
Instead, the unit cell is a cube of volume $(2a)^3$ that contains two of those building blocks, as shown in Fig.~\ref{fig:unitcell}.
This unit cell is also the smallest instance of an ssc lattice for which periodic boundary conditions can be applied.
When sites and links on the outer boundaries are drawn in all of their equivalent boundary locations, the unit cell in Fig.~\ref{fig:unitcell}
can be drawn as Fig.~\ref{fig:ssc2x2}.
Here we consider the basis states of this periodic unit cell with gauge fields truncated to $j_\text{max}=\frac{1}{2}$.

There are 8 unique lattice sites and 12 unique gauge links in the unit cell.
According to the qubit mapping suggested in Sec.~\ref{sec:mapping}, there will be one qubit for each link and one qubit for each site,
which is 20 qubits in total.
It turns out that 215977 of the $2^{20}$ basis states satisfy Gauss's law, which requires a theoretical minimum of 18 qubits, so our mapping to 20
qubits is quite efficient.

To understand the counting of the 215977 physical basis states, it is useful to notice that baryon number $B$ is a conserved quantum number in the SU(2) theory.
Each quark has baryon number $+\frac{1}{2}$ and each antiquark has baryon number $-\frac{1}{2}$.
The Hamiltonian can only create and annihilate fermions in pairs, always one quark with one antiquark, so baryon number remains constant.
This means the Hamiltonian is block diagonal in $B$ and we can count the basis states within each block separately.

Because the unit cell has four quark sites and four antiquark sites, the allowed values for the integer $B$ are $-4\leq B\leq 4$.
For any chosen $B$, it is straightforward to count all possible locations for the quarks and antiquarks, giving the number of
basis states in the absence of any gauge field fluctuations.
For $j_{\rm max}=\frac{1}{2}$, those gauge fluctuations provide a multiplicative factor of 32 as known from the pure gauge studies with a triamond lattice
(which is topologically equivalent to the ssc lattice) \cite{Kavaki:2024ijd}.
Finally, we must recall that every vertex having $j_x=j_y=j_z=j_v=\frac{1}{2}$ corresponds to two physical states ($j_{zv}=0$ or 1) and both of them must be counted.
Taking all of this into account, Table~\ref{tab:statecount} lists the number of basis states in each $B$ sector and
shows that they sum to 215977.

\begin{table}
\caption{The number of Gauss-preserving basis states for a periodic unit cell of the ssc lattice, for $j_{\rm max}=\frac{1}{2}$.
         $B$ denotes baryon number.
         $N_f$ is the number of basis states for the pure fermion theory.
         $N_g$ is the multiplicity coming from gauge fields with $j\in\{0,\frac{1}{2}\}$.
         $N_v$ is the number of additional states arising because there are two options, not just one, at every vertex having $j_x=j_y=j_z=j_v=\frac{1}{2}$.
         The total number of Gauss-preserving basis states is $N_B=N_fN_g+N_v$.
        }\label{tab:statecount}
\begin{ruledtabular}
\begin{tabular}{rrrrr}
$B$   & $N_f$ & $N_g$ &  $N_v$ &  $N_B$ \\
\hline
  $4$ &     1 & 32    &      0 &     32 \\
  $3$ &    36 & 32    &    528 &   1680 \\
  $2$ &   266 & 32    &   7026 &  15538 \\
  $1$ &   784 & 32    &  27120 &  52208 \\
  $0$ &  1107 & 32    &  41637 &  77061 \\
 $-1$ &   784 & 32    &  27120 &  52208 \\
 $-2$ &   266 & 32    &   7026 &  15538 \\
 $-3$ &    36 & 32    &    528 &   1680 \\
 $-4$ &     1 & 32    &      0 &     32 \\
\hline
total &  3281 & 32    & 110985 & 215977
\end{tabular}
\end{ruledtabular}
\end{table}

Physics in the $B=4$ sector has only a single fermionic option because all quark sites are full and all antiquark sites are empty.
From such a state, kinetic terms in the Hamiltonian can neither create nor annihilate a quark-antiquark pair.
Every quark site has a pair of quarks forming a local gauge-singlet baryon, and these baryons are unable to move.
Therefore states in this sector can differ only by their gauge fields.
Similarly, the $B=-4$ sector has four immobile antibaryons.

All other sectors accommodate both pair creation and pair annihilation, and therefore also fermion mobility.
For example, the $B=3$ sector can have six quarks and no antiquarks, or seven quarks plus a single antiquark, or eight quarks plus two antiquarks.
In each case, there are multiple locations available to the quarks and antiquarks, leading to several basis states.
Such states represent three baryons accompanied in some cases by additional mesons or a baryon-antibaryon pair.

The largest sector has $B=0$ and the theory's vacuum state is in this sector.
The bare vacuum is simply the absence of all fermions and gauge excitations, but the interacting vacuum will be a particular superposition
of all 77061 states within the sector.

The ssc unit cell offers a first step into 3D simulations for a non-Abelian gauge theory that includes fermions.
It represents a challenge that could be appropriate for near-term quantum computing hardware.

\section{Outlook}\label{sec:outlook}

In this work, staggered fermions on a semi-simple cubic lattice are found to offer an efficient 3D framework for the quantum computation of non-Abelian gauge theories.
The lattice has only half the gauge links of a standard cubic lattice but still provides 3D derivatives locally at each lattice site.
With only three gauge links at every site, the need to manage partial sums is minimized.
Getting to 3D is crucial for particle physics applications, and hardware-efficient approaches are especially valuable at this early stage,
which makes the ssc lattice an attractive option.

Quantum imaginary time evolution was previously used on an IBM quantum computer to find the ground state of the pure gauge theory
on a topologically equivalent lattice, called the triamond lattice \cite{Kavaki:2024ijd}.
Inclusion of fermions leads naturally to the ssc lattice and, as shown by comparing $N_g$ with the total $N_B$ in Table~\ref{tab:statecount},
dynamical fermions dramatically increase the physical Hilbert space.
This allows the ssc lattice to sustain a full suite of physics phenomenology.

The ten-sided plaquettes of an ssc lattice mean that 3D computations can require significant qubit/qudit connectivity.
Perhaps the ssc lattice will be a good showcase for the various quantum hardware platforms.
Also, while our work has focused on SU(2) gauge theory, the ssc lattice is certainly applicable more broadly than that.

\begin{acknowledgments}
This work was supported in part by the Natural Sciences and Engineering Research Council (NSERC) of Canada.
S.P.\ received additional funding from an Earle Nestmann Undergraduate Research Award.
\end{acknowledgments}

\appendix

\section{Tetrahedral symmetry}\label{app:tetrahedral}

The physics of angular momentum can be obtained from the irreducible representations (irreps) of the rotation group.
In the continuum, these are labeled by the orbital angular momentum quantum number $\ell=0,1,2,3,\ldots$.
A simple cubic lattice retains only a subgroup, namely the octahedral symmetry group, with irreps conventionally named $A_1$,
$A_2$, $E$, $T_1$ and $T_2$.
Group theory can be used to build a subduction table that expresses the octahedral irreps in terms of the continuum $\ell$ irreps
\cite{Johnson:1982yq}.

The triamond lattice has a screw-symmetric version of the octahedral group, but screw translations are not needed when linear momentum is zero
(due to invariance under spatial translations), thus allowing use of the standard octahedral subduction table.
Of interest in the present work is the ssc lattice, which retains a smaller subgroup of the continuum rotations,
specifically a screw-symmetric version of the tetrahedral group.
Because screw translations are not needed at zero linear momentum, the standard tetrahedral group is sufficient here.

\begin{figure}
\includegraphics[width=55mm]{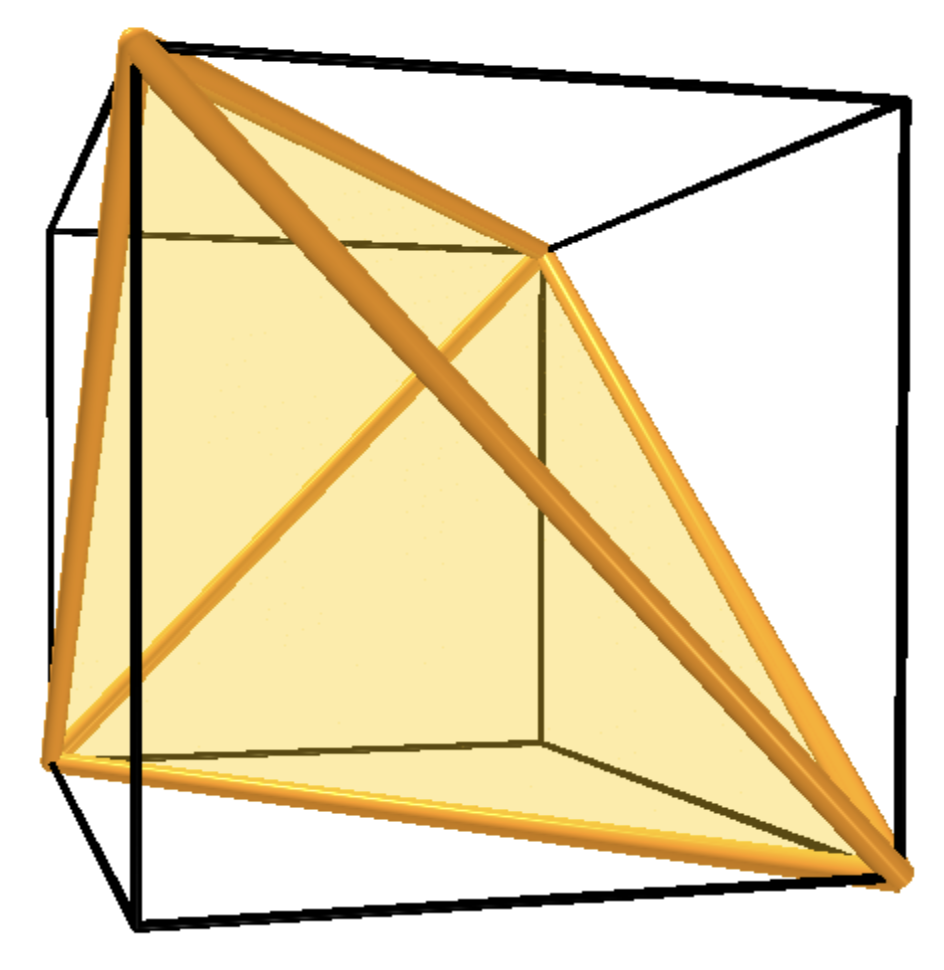}
\caption{A regular tetrahedron inside a cube.
         The tetrahedron is invariant under a subgroup of the rotations that leave the cube invariant.
}\label{fig:tetracube}
\end{figure}

To understand the connection between an ssc lattice and tetrahedral symmetry, consider Fig.~\ref{fig:tetracube}.
When the tetrahedron is absent, the cube has 4-fold, 3-fold and 2-fold symmetry axes piercing its faces, corners and edges respectively.
The embedded tetrahedron retains the 3-fold symmetries but keeps only 2-fold rotations at the cube faces.
These tetrahedral invariances are precisely the symmetries of the ssc lattice, aside from screw translations.

The irreps of the tetrahedral group are named $A$, $E_+$, $E_-$ and $T$.
We connect these to continuum angular momentum by following the approach used for the octahedral group in Ref.~\cite{Johnson:1982yq}, as follows.

The tetrahedral group has 12 elements in four conjugacy classes.
Class $E$ contains only the identity.
Class $C_2$ contains three elements, which are $\pi$ rotations about the three 2-fold symmetry axes.
Class $C_3$ contains four elements, which are $2\pi/3$ rotations around the four 3-fold symmetry axes.
Class $C_3^\prime$ is the same as $C_3$ but for rotations of $4\pi/3$.
The connection between conjugacy classes and irreps is given by the character table of Table~\ref{tab:characters}.
Having obtained the characters, the next step is to calculate multiplicities that project the tetrahedral group
onto the continuum group.
Results are given in Table~\ref{tab:angmom}.
\begin{table}
\caption{A character table for the tetrahedral group.
         Each column is a conjugacy class and each row is an irreducible representation.
        }\label{tab:characters}
\begin{ruledtabular}
\begin{tabular}{lcccc}
      & $E$ & $C_2$ & $C_3$           & $C_3^\prime$    \\
\hline
$A$   & 1   & 1     & 1               & 1               \\
$E_+$ & 1   & 1     & $e^{2\pi i/3}$  & $e^{-2\pi i/3}$ \\
$E_-$ & 1   & 1     & $e^{-2\pi i/3}$ & $e^{2\pi i/3}$  \\
$T$   & 3   & $-1$  & 0               & 0
\end{tabular}
\end{ruledtabular}
\end{table}

Table~\ref{tab:angmom} shows that $\ell=0$ comes from the one-dimensional irrep $A$.
For the ssc lattice, this scalar irrep is the sum of all plaquettes that we already obtained in Eq.~(\ref{eq:scalar}).
Table~\ref{tab:angmom} also shows that $\ell=1$ comes from the three-dimensional irrep $T$.
For the ssc lattice, this vector irrep is the set of three differences among plaquettes that we obtained
in Eqs.~(\ref{eq:vector1}-\ref{eq:vector3}).

Table~\ref{tab:angmom} contains the information needed for higher $\ell$ values as well.
For example, $\ell=2$ is seen to arise from the three components in $T$ and from the two components in $E_\pm$.
The $E_\pm$ irreps of an ssc lattice can be constructed from differences between the following three operators
\begin{align}
\sum_{\vec n}\left( P_{\bar r}(\vec n) + P_{\bar c}(\vec n) \right) \,, \label{eq:E1} \\
\sum_{\vec n}\left( P_{\bar g}(\vec n) + P_{\bar m}(\vec n) \right) \,, \label{eq:E2} \\
\sum_{\vec n}\left( P_{\bar b}(\vec n) + P_{\bar y}(\vec n) \right) \,. \label{eq:E3}
\end{align}
Three differences can be formed (\ref{eq:E1} minus \ref{eq:E2}, \ref{eq:E1} minus \ref{eq:E3} and \ref{eq:E2} minus \ref{eq:E3})
but only two are linearly independent, corresponding to $E_+$ and $E_-$.
\begin{table}
\caption{Number of copies of each tetrahedral irrep ($A,E_\pm,T$) in the subduced representation of continuum angular momentum (quantum number $\ell=0, 1, 2, \ldots$).
        }\label{tab:angmom}
\begin{ruledtabular}
\begin{tabular}{ccccccccc}
$\ell$& 0 & 1 & 2 & 3 & 4 & 5 & 6 & 7 \\
\hline
$A$   & 1 & 0 & 0 & 1 & 1 & 0 & 2 & 1 \\
$E_+$ & 0 & 0 & 1 & 0 & 1 & 1 & 1 & 1 \\
$E_-$ & 0 & 0 & 1 & 0 & 1 & 1 & 1 & 1 \\
$T$   & 0 & 1 & 1 & 2 & 2 & 3 & 3 & 4
\end{tabular}
\end{ruledtabular}
\end{table}

In summary, the change from a simple cubic lattice to the ssc lattice results in a rather minor adjustment for angular momentum.
The simple cubic lattice has five irreps and the ssc lattice has four.

\section{Fermion/antifermion lattice sites}\label{app:fermionsites}

At zero momentum, the Dirac Hamiltonian reduces to $m\Psi^\dagger\gamma^0\Psi$.
Therefore the $+1$ and $-1$  eigenstates of $\gamma^0$ correspond to particles and antiparticles respectively,
and the projectors $P_\pm=\frac{1}{2}(1\pm\gamma^0)$ isolate the particle and antiparticle components.
In this appendix, we show that these particle and antiparticle components live on alternating (even versus odd)
sites of the staggered lattice.

Begin by decomposing the full fermion field $\Psi(\vec n)$ into eigenstates of $\gamma^0$,
\begin{equation}
\Psi(\vec n) = \Psi_+(\vec n) + \Psi_-(\vec n) \,,
\end{equation}
where
\begin{equation}
\Psi_\pm(\vec n) = P_\pm\Psi(\vec n) \,.
\end{equation}
Now recall the transformation used in Eq.~(\ref{eq:alpha}),
\begin{equation}
\Psi(\vec n) = \alpha(\vec n)\chi(\vec n) \,,
\end{equation}
where
\begin{equation}
\alpha(\vec n) = (i\gamma^0\gamma^1)^{n_x}(i\gamma^0\gamma^2)^{n_y}(i\gamma^0\gamma^3)^{n_z} \,.
\end{equation}
This will allow $\Psi_\pm(\vec n)$ to be expressed in terms of $\chi(\vec n)$ and its components, $\chi_\pm(\vec n)=P_\pm\chi(\vec n)$.
To do so, commutation relations among the Dirac gamma matrices lead to a useful expression,
\begin{equation}
P_\pm\alpha(\vec n) = \alpha(\vec n)\frac{1}{2}(1\pm\epsilon(\vec n)\gamma^0) \,,
\end{equation}
where the site parity is
\begin{equation}
\epsilon(\vec n) = (-1)^{n_x+n_y+n_z} = \left\{\begin{array}{rl} 1 & ~\text{for even}~ \vec n \,, \\
                                                                -1 & ~\text{for odd}~ \vec n \,. \end{array}\right.
\end{equation}
This means the original fermion field can be written as follows,
\begin{eqnarray}
\Psi_\pm(\vec n) &=& P_\pm\alpha(\vec n)\chi(\vec n) \\
                 &=& \alpha(\vec n)\frac{1}{2}(1\pm\epsilon(\vec n)\gamma^0)\chi(\vec n) \\
                 &=& \alpha(\vec n)\frac{1}{2}(1\pm\epsilon(\vec n)\gamma^0)\left(\chi_+(\vec n)+\chi_-(\vec n)\right) \\
                 &=& \alpha(\vec n)\frac{1}{2}\bigg((1\pm\epsilon(\vec n))\chi_+(\vec n)+(1\mp\epsilon(\vec n))\chi_-(\vec n)\bigg) \nonumber \\
\end{eqnarray}
Projectors for even and odd sites have emerged in that last line,
\begin{eqnarray}
P_e(\vec n) &=& \frac{1}{2}(1+\epsilon(\vec n)) \,, \\
P_o(\vec n) &=& \frac{1}{2}(1-\epsilon(\vec n)) \,.
\end{eqnarray}
Recall from Sec.~\ref{sec:fermions} that we are omitting $\chi_-$ and keeping only $\chi_+$ (and naming it simply $\chi$),
so we have arrived at
\begin{align}
\Psi_+(\vec n) = \alpha(\vec n)\chi(\vec n) & ~\text{for even}~ \vec n, \\
\Psi_-(\vec n) = \alpha(\vec n)\chi(\vec n) & ~\text{for odd}~ \vec n.
\end{align}
Thus, the $\Psi_+$ and $\Psi_-$ degrees of freedom get mapped to the even and odd sites respectively in the staggered formalism.

\section{More about the kinetic terms}\label{app:morekinetic}

The $z$ component of the Hamiltonian's kinetic terms is given in Eq.~(\ref{eq:Hkz}).
The $x$ and $y$ components are similar but they need to be translated into the $z$ basis.
This appendix presents that calculation.

Consider a lattice site where the $x$ and $y$ gauge links each have $j=\frac{1}{2}$, but the $z$ link has $j=0$ and no fermion is present.
Gauss's law requires that the $x$ and $y$ links form a gauge singlet at the site.
If we now apply a kinetic term along the $z$ link, it will flip that link to $j=\frac{1}{2}$ and will also create a quark at the site.
Gauss's law requires that the $z$ link and the fermion form a gauge singlet at the site.
Therefore the state of our system at that site is
\begin{eqnarray}
\left|S_{xy,zv}\right> &=& \frac{1}{\sqrt{2}}\left|\uparrow_x\downarrow_y-\downarrow_x\uparrow_y\right>
                           \frac{1}{\sqrt{2}}\left|\uparrow_z\downarrow_v-\downarrow_z\uparrow_v\right> ~~~~~ \\
                       &=& \left|0,0\right>_{xy}\left|0,0\right>_{zv} \\
                       &=& \left|0,0\right>_{0\otimes0}
\end{eqnarray}
where the subscript $v$ denotes the vertex qubit representing the quark.
This is already in our chosen computational basis, where the qubit pairs are $xy$ and $zv$.

Alternatively, suppose the $y$ and $z$ gauge links initially formed a singlet, and we then applied the kinetic operator along the $x$ link.
The result is
\begin{equation}
\left|S_{yz,xv}\right> = \frac{1}{\sqrt{2}}\left|\uparrow_y\downarrow_z-\downarrow_y\uparrow_z\right>
                         \frac{1}{\sqrt{2}}\left|\uparrow_x\downarrow_v-\downarrow_x\uparrow_v\right>
\end{equation}
but we need to translate this result into  our computational basis:
\begin{eqnarray}
\left|S_{yz,xv}\right>
&=& \frac{1}{2}\big|\uparrow_x\uparrow_y\downarrow_z\downarrow_v-\downarrow_x\uparrow_y\downarrow_z\uparrow_v
    -\uparrow_x\downarrow_y\uparrow_z\downarrow_v \nonumber \\
& & +\downarrow_x\downarrow_y\uparrow_z\uparrow_v\big> \\
&=& \frac{1}{2}\left|1,1\right>_{xy}\left|1,-1\right>_{zv} - \frac{1}{2}\left|1,0\right>_{xy}\left|1,0\right>_{zv} \nonumber \\
& & - \frac{1}{2}\left|0,0\right>_{xy}\left|0,0\right>_{zv} + \frac{1}{2}\left|1,-1\right>_{xy}\left|1,1\right>_{zv} \hspace{8mm} \\
&=& \frac{\sqrt{3}}{2}\left|0,0\right>_{1\otimes1} - \frac{1}{2}\left|0,0\right>_{0\otimes0} \,. \label{eq:xv}
\end{eqnarray}
A similar calculation gives the third option, which is the kinetic operator along the $y$ link,
\begin{eqnarray}
\left|S_{zx,yv}\right> &=& \frac{1}{\sqrt{2}}\left|\uparrow_z\downarrow_x-\downarrow_z\uparrow_x\right>
                           \frac{1}{\sqrt{2}}\left|\uparrow_y\downarrow_v-\downarrow_y\uparrow_v\right> \hspace{8mm} \\
                       &=& -\frac{\sqrt{3}}{2}\left|0,0\right>_{1\otimes1} - \frac{1}{2}\left|0,0\right>_{0\otimes0} \,. \label{eq:yv}
\end{eqnarray}
The three cases are related by
\begin{equation}
\left|S_{xy,zv}\right>+\left|S_{yz,xv}\right>+\left|S_{zx,yv}\right>=0 \,.
\end{equation}

To implement the kinetic Hamiltonian in terms of Pauli gates, it is convenient to begin with a table of required qubit actions.
Table~\ref{tab:Hkz} shows the effect of $H_k$ terms along the $z$ direction.
There are only eight rows in this table because the other eight basis states are either unphysical or cannot be raised/lowered further due
to having the extremal fermion number.
Transforming this table into Pauli gates results in Eq.~(\ref{eq:Hkz}).
\begin{table}
\caption{Action of the Hamiltonian's kinetic terms along the $z$ axis.
         One term raises the quark number and the other lowers it.
         Notation is $\left|q_xq_yq_zq_v\right>$.
        }\label{tab:Hkz}
\begin{ruledtabular}
\begin{tabular}{cc}
raising & lowering \\
\hline
$\left|0000\right>\to\left|0010\right>$ & $\left|0001\right>\to\left|0010\right>$ \\
$\left|1100\right>\to\left|1110\right>$ & $\left|1101\right>\to\left|1110\right>$ \\
$\left|1010\right>\to\left|1000\right>$ & $\left|1011\right>\to\left|1000\right>$ \\
$\left|0110\right>\to\left|0100\right>$ & $\left|0111\right>\to\left|0100\right>$ \\
\hline
$\left|1000\right>\to\left|1011\right>$ & $\left|1000\right>\to\left|1010\right>$ \\
$\left|0100\right>\to\left|0111\right>$ & $\left|0100\right>\to\left|0110\right>$ \\
$\left|0010\right>\to\left|0001\right>$ & $\left|0010\right>\to\left|0000\right>$ \\
$\left|1110\right>\to\left|1101\right>$ & $\left|1110\right>\to\left|1100\right>$
\end{tabular}
\end{ruledtabular}
\end{table}

\begin{table}
\caption{Action of the Hamiltonian's kinetic terms along the $y$ axis.
         One term raises the quark number and the other lowers it.
         Notation is $\left|q_xq_yq_zq_v\right>$ and $\left|111C\right>=-\frac{1}{2}\left|1110\right>-\frac{\sqrt{3}}{2}\left|1111\right>$.
        }\label{tab:Hky}
\begin{ruledtabular}
\begin{tabular}{cc}
raising & lowering \\
\hline
$\left|0000\right>\to\left|0100\right>$ & $\left|0001\right>\to\left|0100\right>$ \\
$\left|1100\right>\to\left|1000\right>$ & $\left|1101\right>\to\left|1000\right>$ \\
$\left|0110\right>\to\left|0010\right>$ & $\left|0111\right>\to\left|0010\right>$ \\
$\left|1010\right>\to\left|111C\right>$ & $\left|1011\right>\to\left|111C\right>$ \\
\hline
$\left|1000\right>\to\left|1101\right>$ & $\left|1000\right>\to\left|1100\right>$ \\
$\left|0100\right>\to\left|0001\right>$ & $\left|0100\right>\to\left|0000\right>$ \\
$\left|0010\right>\to\left|0111\right>$ & $\left|0010\right>\to\left|0110\right>$ \\
$\left|111C\right>\to\left|1011\right>$ & $\left|111C\right>\to\left|1010\right>$
\end{tabular}
\end{ruledtabular}
\end{table}
Table~\ref{tab:Hky} shows the effect of $H_k$ terms along the $y$ direction.
Once again the physical transitions fit into eight rows, but now we see the presence of the state $\left|111C\right>$
because of the translation into the computational basis.
The kinetic terms in the $y$ direction are
\begin{eqnarray}
\left(H_k\right)_y &=& \sum_{y~{\rm links}}(-1)^{n_x}\tilde D_{\vec n}^\dagger X_y\tilde D_{\vec n+\hat y}
\end{eqnarray}
and the expression for $\tilde D$ comes from Table~\ref{tab:Hky}.
Specifically,
\begin{eqnarray}
\tilde D &=& \left(\Pi_y^+X_v\Pi_v^- + \Pi_y^-\Pi_v^+\right)\Pi_x^+\Pi_z^+ \nonumber \\
         & & + \left(\Pi_y^-X_v\Pi_v^- + \Pi_y^+\Pi_v^+\right)\left(\Pi_x^-\Pi_z^+ + \Pi_x^+\Pi_z^-\right) \nonumber \\
         & & + \Pi_y^-\left(-\tfrac{1}{2}\Pi_v^+-\tfrac{\sqrt{3}}{2}X_v\Pi_v^-\right)\Pi_x^-\Pi_z^- \nonumber \\
         & & + \Pi_y^+\left(-\tfrac{1}{2}X_v\Pi_v^--\tfrac{\sqrt{3}}{2}\Pi_v^-\right)\Pi_x^-\Pi_z^- \,,
\end{eqnarray}
where
\begin{equation}
\Pi_j^\pm = \tfrac{1}{2}(1\pm Z_j) \,.
\end{equation}
The kinetic Hamiltonian in the $x$ direction differs from the $y$ direction in three ways:
interchange of the $x$ and $y$ qubits, a change of sign for $\sqrt{3}$ because of Eqs.~(\ref{eq:xv}) and (\ref{eq:yv}),
and removal of the staggered phase $(-1)^{n_x}$ due to Eq.~(\ref{eq:Hk}).

\end{document}